\documentclass[10pt,twocolumn]{IEEEtran}%
\usepackage{graphicx}
\usepackage{amsmath}%
\usepackage{amsfonts}%
\usepackage{amssymb}
\makeatletter
\def\ifundefined{\@ifundefined}
\makeatother
\let\OLDitemize\itemize\let\OLDenumerate\enumerate\let\OLDdescription\description
\renewcommand{\itemize}[1][\relax]{\OLDitemize} \renewcommand{\enumerate
}[1][\relax]{\OLDenumerate} \renewcommand{\description}[1][\relax
]{\OLDdescription}
\providecommand{\pubid}[1]{\relax}

\providecommand{\specialpapernotice}[1]{\relax} 
\let\CMPARstart\PARstart
\let\OLDappendix\appendix\renewcommand{\appendix}[1][\relax]{\OLDappendix}
\newif\ifuseRomanappendices\useRomanappendicestrue
\let\OLDbiography\biography\let\OLDendbiography\endbiography\renewcommand
{\biography}[2][\relax]{\OLDbiography{#2}}
\renewcommand{\endbiography }{\OLDendbiography}
\renewcommand{\PARstart}[2]{\CMPARstart{#1}{#2}}
\newcounter{num}

\begin{document}

\title{Long Nonbinary Codes Exceeding the Gilbert-Varshamov bound for Any Fixed Distance}
\author{Sergey Yekhanin \qquad Ilya Dumer \thanks{ S. Yekhanin is a Ph.D. student at
the Department of Electrical Engineering and Computer Science, Massachusetts
Institute of Technology, Cambridge, MA 02139, USA. His research was supported
in part by NTT Award MIT 2001-04 and NSF grant CCR 0219218. I.\ Dumer is with
the College of Engineering, University of California, Riverside, CA 92521,
USA. His research was supported by NSF grant CCR 0097125.}}
\maketitle

\begin{abstract}
Let $A(q,n,d)$ denote the maximum size of a $q$-ary code of length $n$ and
distance $d$. We study the minimum asymptotic redundancy $\rho(q,n,d)=n-\log
_{q}A(q,n,d)$ as $n$ grows while $q$ and $d$ are fixed. For any $d$ and $q\geq
d-1,$ long algebraic codes are designed that improve on the BCH codes and have
the lowest asymptotic redundancy
\[
\rho(q,n,d)\lesssim((d-3)+1/(d-2))\log_{q}n
\]
known to date. Prior to this work, codes of fixed distance that asymptotically
surpass BCH codes and the Gilbert-Varshamov bound were designed only for
distances $4,5,$~and~$6$.

\end{abstract}



\section{Introduction}

\PARstart{L}{et} $A(q,n,d)$ denote the maximum size of a $q$-ary code of
length $n$ and distance $d$. We study the asymptotic size $A(q,n,d)$ if $q$
and $d$ are fixed as $n\rightarrow\infty$, and introduce a related quantity
\[
c(q,d)={{\underline{\lim}}_{n\rightarrow\infty}}\frac{n-\log_{q}{A(q,n,d)}%
}{\log_{q}n},
\]
which we call the \textit{redundancy coefficient. }

\smallskip

The Hamming upper bound
\[
A(q,n,d)\leq q^{n}\left/  \sum_{i=0}^{\lfloor(d-1)/2\rfloor}(q-1)^{i}%
{\binom{n}{i}}\right.
\]
\ leads to the lower bound
\begin{equation}
c(q,d)\geq\left\lfloor (d-1)/2\right\rfloor , \label{Hamming_bound}%
\end{equation}
which is the best bound on $c(q,d)$ known to date for arbitrary values of $q$
and $d$. On the other hand, the Varshamov existence bound admits any linear
$[n,k,d]_{q}$ code of dimension
\[
k\leq n-1-\left\lfloor \log_{q}\sum_{i=0}^{d-2}(q-1)^{i}{\binom{n-1}{i}%
}\right\rfloor .
\]
This leads to the redundancy coefficient
\begin{equation}
c(q,d)\leq d-2. \label{Varshamov_bound}%
\end{equation}
(Note that the Gilbert bound results in a weaker inequality $c(q,d)\leq d-1$.)

Let $e$ be a primitive element of the Galois field $F_{q^{m}}.$ Consider
(see~\cite{MS}) the narrow-sense BCH code defined by the generator polynomial
with \textit{zeros} $e^{1},...,e^{d-2}$. Let $C_{q}^{m}(d)$ denote the
extended BCH code obtained by adding the overall parity check. Code $C_{q}%
^{m}(d)$ has length $\ q^{m},$ \textit{constructive }distance $d$ , and
redundancy coefficient
\begin{equation}
c(q,d)\leq\left\lceil \frac{(d-2)(q-1)}{q}\right\rceil . \label{BCH_bound}%
\end{equation}
Note that the above BCH bound~(\ref{BCH_bound}) is better than the Varshamov
bound~(\ref{Varshamov_bound}) for $q<d-1$ and coincides
with~(\ref{Varshamov_bound}) for $q\geq d-1$. Note also that~(\ref{BCH_bound})
meets the Hamming bound~(\ref{Hamming_bound}) if $q=2$ or $d=3$. Therefore
\[
c(2,d)=\left\lfloor (d-1)/2\right\rfloor \mbox{ and }c(q,3)=1.
\]

For distances $4,5,$ and $6,$ infinite families of nonbinary linear codes are
constructed in \cite{Dumer_88} and ~\cite{Dumer_95} that reduce asymptotic
redundancy (\ref{BCH_bound}). Open Problem 2 from \cite{Dumer_95} also raises
the question if the BCH bound~(\ref{BCH_bound}) can be improved for larger
values of $d$. Our main result is an algebraic construction of codes that
gives an affirmative answer to this problem for all $q\geq d-1$. In terms of
redundancy, the new bound is expressed by

\textbf{Theorem \refstepcounter{num}\arabic{num} \label{Main_theorem}:} For
all $q$ and $d\geq3,$
\begin{equation}
c(q,d)\leq(d-3)+1/(d-2). \label{New_bound}%
\end{equation}
\smallskip Combining~(\ref{BCH_bound}) and~(\ref{New_bound}), we obtain
\[
c(q,d)\leq\min\left(  \left\lceil \frac{(d-2)(q-1)}{q}\right\rceil
,(d-3)+\frac{1}{(d-2)}\right)  ,
\]
Note that the above bound is better than the Varshamov existence bound for
arbitrary $q$ and $d\geq4$.

The rest of the paper is organized as follows. In Section II, we review the
upper bounds for $c(q,d)$ that surpass the BCH bound~(\ref{BCH_bound}) for
small values of $d$. In Section III, we present our code construction and
prove the new bound~(\ref{New_bound}). This proof rests on important
Theorem~\ref{Lines}, which is proven in Section IV. Finally, we make some
concluding remarks in section V.

\section{Previous work}

Prior to this work, codes that asymptotically exceed the BCH
bound~(\ref{BCH_bound}) were known only for $d\leq6$. We start with the bounds
for $c(q,4)$. Linear $[n,n-\rho,4]_{q}$ codes are equivalent to \textit{caps}
in projective geometries $PG(\rho-1,q)$ and have been studied extensively
under this name. See~\cite{Hirshfeld} for a review. However, the exact values
of $c(q,4)$ remain unknown for all $q\geq3,$ and the gaps between the upper
and the lower bounds are still large.

The Hamming bound yields $c(q,4)\geq1$. Mukhopadhyay~\cite{Mukhopadhyay}
obtained the upper bound $c(q,4)\leq1.5$. For all values of $q,$ this was
later improved by Edel and Bierbrauer~\cite{Edel_d4} to
\begin{equation}
c(q,4)\leq\frac{6}{\log_{q}{(q^{4}+q^{2}-1)}}. \label{d4}%
\end{equation}
Note that for large values of $q$ the right hand side of~(\ref{d4}) tends
to~$1.5$. The case of $q=3$ has been of special interest, and general bound
(\ref{d4}) has been improved in a few papers (see~\cite{Hill}, \cite{Frankl},
\cite{Calderbank}, \cite{Edel_d4_q3}). The current record
\[
c(3,4)\leq1.3796
\]
due to Edel~\cite{Edel_d4_q3} slightly improves on the previous record
$c(3,4)\leq1.3855$ obtained by Calderbank and Fishburn~\cite{Calderbank}. For
$q=4,$ the construction of ~\cite{Glynn} also improves~(\ref{d4}). Namely,
$c(4,4)\leq1.45$.

\smallskip

Now we proceed to the bounds for $c(q,5)$. The Hamming bound yields
$c(q,5)\geq2$. Several families of linear codes constructed in \cite{Dumer_95}
reach the bound
\begin{equation}
c(q,5)\leq7/3 \label{d5}%
\end{equation}
for all values of $q$. Later, alternative constructions of codes with the same
asymptotic redundancy were considered in~\cite{Feng}. \ Similarly to the case
of $d=4,$ there exist better bounds for small alphabets. Namely, Goppa pointed
out that ternary double error-correcting BCH codes asymptotically meet the
Hamming bound~(\ref{Hamming_bound}). For $q=4$ and $d=5,$ two different
constructions that asymptotically meet the Hamming bound were proposed
in~\cite{Gevorkyan_75} and~\cite{Dumer_Zinoviev}. Thus,
\[
c(3,5)=c(4,5)=2.
\]

\smallskip

For $d=6,$ the infinite families of linear codes designed in \cite{Dumer_88}
and \cite{Dumer_95} reach the upper bound
\begin{equation}
c(q,6)\leq3 \label{d6}%
\end{equation}
for all $q$. The constructions are rather complex and the resulting linear
codes are not cyclic. Later, a simpler construction of a cyclic code with the
same asymptotic redundancy was proposed in~\cite{Danev}. Again, better bounds
exist for small values of $q$. Namely, $c(3,6)\leq2.5$~\cite{Dumer_95} and
$c(4,6)\leq17/6$~\cite{Feng_d6}.\smallskip

We summarize the bounds described so far in Figure~1. \begin{figure}[ptb]
{\includegraphics[scale=0.5]{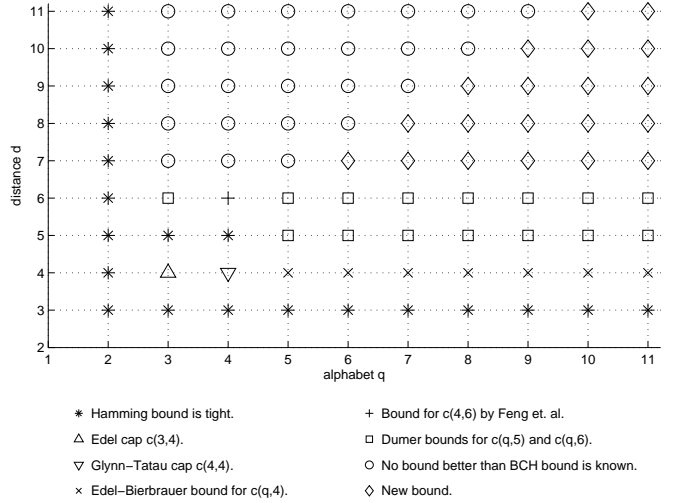}}\caption{A taxonomy of best known
upper bounds for $c(q,d)$.}%
\label{Bounds_picture}%
\end{figure}

\smallskip

The following Lemma~\ref{Monotonicity} due to Gevorkyan~\cite{Gevorkyan_79}
shows that redundancy $c(q,d)$ cannot increase when the alphabet size is reduced.

\textbf{Lemma \refstepcounter{num}\arabic{num} \label{Monotonicity}:} For
arbitrary value of distance $d$,
\[
q_{1}\leq q_{2} \ \Rightarrow\ c(q_{1},d)\leq c(q_{2},d).
\]

\textbf{Proof:} Given a code $V$ of length $n$ over the $q_{2}$-ary alphabet
we prove the existence of a code $V^{\prime}$ of the same length over $q_{1}%
$-ary alphabet with the same redundancy coefficient. Let $q_{2}$-ary alphabet
be an additive group $E_{q_{2}}$, and $q_{1}$-ary alphabet form a subset
$E_{q_{1}}\subseteq E_{q_{2}}$. Define the componentwise shift $V_{v}=V+v$ of
code $V$ by an arbitrary vector $v\in E_{q_{2}}^{n}$. Note that any vector
$f\in E_{q_{1}}^{n}$ belongs to exactly $|V|$ codes among all $q_{2}^{n}$
codes $V_{v}$, as $v$ runs through $E_{q_{2}}^{n}$. Hence, codes $V_{v}$
include on average $q_{1}^{n}|V|/q_{2}^{n}$ vectors of the subset $E_{q_{1}%
}^{n}\subseteq E_{q_{2}}^{n}$. Therefore, some set $V_{v}\cap E_{q_{1}}^{n}$
has at least this average size. Denote this set by $V^{\prime}$. Clearly,
$V^{\prime}$ is a $q_{1}$-ary code with the same distance as code $V$. It
remains to note that
\[
\frac{n-\log_{q_{1}}\left(  q_{1}^{n}|V|/q_{2}^{n}\right)  }{\log_{q_{1}}%
n}=\frac{n-\log_{q_{2}}|V|}{\log_{q_{2}}n}.
\]
The proof is completed. {\hfill$\Box$}

\textbf{Corollary \refstepcounter{num}\arabic{num} \label{Alphabet_decrease}:}
Let $\{q_{i}\}$ be an infinite sequence of growing alphabet sizes. Assume
there exist $c^{\star}$ and $d$ such that for all $i$, $c(q_{i},d)\leq
c^{\star}$. Then $c(q,d)\leq c^{\star}$ for all values of $q$.

\textbf{Proof:} This follows trivially from Lemma~\ref{Monotonicity}.
{\hfill$\Box$}

\section{Code Construction}

In the sequel, the elements of the field $F_{q}$ are denoted by Greek letters,
while the elements of extension fields $F_{q^{i}}$ are denoted by Latin letters.

We start with an extended BCH code $C=C_{q}^{m}(d-1)$ of length $n=q^{m}$ and
constructive distance $d-1.$ Here for any position $j\in\lbrack1,q^{m}],$ we
define its \textit{locator} $e_{j}$, where $e_{j}=e^{j}$ for $j<n$ and
$e_{n}=0.$ Then the parity check matrix of code $C$ has the form
\begin{equation}
H_{q}^{m}(d-1)=\left(
\begin{array}
[c]{lcl}%
1 & \ldots & 1\\
e_{1} & \ldots & e_{n-1}\\
\vdots & \ldots & \vdots\\
e_{1}^{d-3} & \ldots & e_{n-1}^{d-3}%
\end{array}%
\begin{array}
[c]{l}%
1\\
0\\
\vdots\\
0
\end{array}
\right)  . \label{BCH_matrix}%
\end{equation}
Here the powers of locators $e_{j}$ are represented with respect to some basis
of $F_{q^{m}}$ over $F_{q}$. Note that the redundancy of $C$ is at most
$(d-3)m+1$. Also, we assume in the sequel that $q$ does not divide $d-2,$
since code $C$ has constructive distance $d$ instead of $d-1$ otherwise$.$

Consider any nonzero codeword $c\in C$ of weight $w$ with nonzero symbols in
positions $j_{1},...,j_{w}.$ Let $X(c)=\{x_{1},\ldots,x_{w}\}$ denote its
\textit{locator set, }where we use notation $x_{i}=e_{j_{i}}$ for all
$i=1,...,w$. We say that $X(c)$ lies on an affine line $L(a,b)$ over $F_{q}$
if there exist $a,b\in F_{q^{m}}$ such that
\begin{equation}
x_{i}=a+\lambda_{i}b \label{locator_line}%
\end{equation}
where $\lambda_{i}\in F_{q}$ for all values of $i=1,...,w$.

The key observation underlining our code construction is that under some
restrictions on extension $m$ and characteristic char$F_{q}$ of the field
$F_{q},$ any code vector $c\in C$ of weight $d-1$ has its locator set $X(c)$
lying on some affine line.\footnote{We shall also see that code $C_{q}%
^{m}(d-1)$ does have minimum distance $d-1$ under these restrictions.}
Formally, this is expressed by\smallskip

\textbf{Theorem \refstepcounter{num}\arabic{num} \label{Lines}:} Let $m$ be a
prime, $m>(d-3)!$ and char$F_{q}>d-3.$ Consider the extended BCH code
$C_{q}^{m}(d-1)$ of constructive distance $d-1.$ Then any codeword $c$ of
minimum weight $d-1$ has its locator set $X(c)$ lying on some affine line
$L(a,b)$ over~$F_{q}$.\smallskip

We defer the proof of Theorem~\ref{Lines} till section~IV and proceed with the
code construction. Let%
\[
s=\left\lceil m/(d-2)\right\rceil ,\quad\mu=s(d-2).
\]
Consider the field $F_{q^{\mu}}$ and its subfield $F_{q^{s}}$. Let
$g=\{g_{1},\ldots,g_{\mu}\}$ be the basis of $F_{q^{\mu}}$ over $F_{q}$ such
that $F_{q^{s}}$ is spanned by $\{g_{1},\ldots,g_{s}\}$. Let $h=\{h_{1}%
,\ldots,h_{m}\}$ be an arbitrary basis of $F_{q^{m}}$ over $F_{q}$. Below we
map each element $x=$ $\sum\limits_{i=1}^{m}\alpha_{i}h_{i}$ of the field
$F_{q^{m}}$ onto the element
\begin{equation}
\hat{x}=\sum\limits_{i=1}^{m}\alpha_{i}g_{i} \label{Inject}%
\end{equation}
of the field $F_{q^{\mu}}.$ It is readily seen that for arbitrary $a,b\in
F_{q^{m}}$ and $\lambda\in F_{q}$
\begin{equation}
\widehat{a+\lambda b}=\hat{a}+\lambda\hat{b}. \label{Inject_prop}%
\end{equation}
Recall that the \textit{norm}~\cite{FF} of $\hat{x}\in F_{q^{\mu}},$
\begin{equation}
N_{F_{q^{\mu}}\left/  F_{q^{s}}\right.  }(\hat{x})=N_{d-2}(\hat{x})=\hat
{x}^{q^{(d-3)s}+\ldots+q^{s}+1} \label{Norm}%
\end{equation}
is a classical mapping from $F_{q^{\mu}}$ to $F_{q^{s}}$.

\smallskip

Now we are ready to present our code construction. Consider the $q$-ary code
$C^{\prime}(n,k^{\prime},d^{\prime})$ of length $n=q^{m}$ with the parity
check matrix
\begin{equation}
\hat{H}_{q}^{m}=\left(
\begin{array}
[c]{lcll}%
1 & \ldots & 1 & 1\\
e_{1} & \ldots & e_{n-1} & 0\\
\vdots & \ldots & \vdots & \vdots\\
e_{1}^{d-3} & \ldots & e_{n-1}^{d-3} & 0 \smallskip\\
N_{d-2}(\hat{e}_{1}) & \ldots & N_{d-2}(\hat{e}_{n-1}) & 0
\end{array}
\right)  \label{Final_matrix}%
\end{equation}
where the locators $e_{j}$ and their powers are represented in $F_{q}$ with
respect to the basis $h$ and values of $N_{d-2}$ are represented in $F_{q}$
with respect to $g$. Recall that $N_{d-2}(x)$ takes values in $F_{q^{s}}$.
Therefore the redundancy of $C^{\prime}$ does not exceed $(d-3)m+s+1.$

Below is the main theorem of the paper.\smallskip

\textbf{Theorem \refstepcounter{num}\arabic{num} \label{Code_family}:} Suppose
$m>(d-3)!$ is a prime, and char$F_{q}>d-3$; then code $C^{\prime}(n,k^{\prime
},d^{\prime})$ defined by~(\ref{Final_matrix}) has parameters
\[
\lbrack q^{m},k^{\prime}\geq q^{m}-(d-3)m-\left\lceil m/(d-2)\right\rceil
-1,d^{\prime}\geq d]_{q}.
\]

\textbf{Proof:} Note that $d^{\prime}\geq d-1$, since $C^{\prime}$ is a
subcode of the extended BCH code $C$ defined in~(\ref{BCH_matrix}). Let
$C_{d-1}\subseteq C$ be the set of all codewords of weight exactly $d-1$. It
remains to prove that $C^{\prime}\cap C_{d-1}=\emptyset.$

Assume the converse. Let $c\in C^{\prime}$ be a codeword of weight $d-1$ with
locator set $X(c)=(x_{1},\ldots,x_{d-1})$. This implies that for some nonzero
symbols $\xi_{1},\ldots,\xi_{d-1}\in F_{q}:$
\begin{equation}
\left\{
\begin{array}
[c]{lcc}%
\sum\limits_{i=1}^{d-1}\xi_{i}x_{i}^{t} & =0, & t=0,...,d-3;\\
\sum\limits_{i=1}^{d-1}\xi_{i}N_{d-2}(\hat{x}_{i}) & =0. &
\end{array}
\right.  \label{main_system0}%
\end{equation}
Note that $c\in C_{d-1}$. Therefore according to Theorem~\ref{Lines}, there
exist $a,b\neq0$ from $F_{q^{m}}$ and pairwise distinct $\{\lambda_{i}\}\in
F_{q}$ such that $x_{i}=a+\lambda_{i}b.$ Consider the affine permutation
$\pi(x)=A+Bx$ of the entire locator set $F_{q^{m}}$, where $A=-ab^{-1}$ and
$B=b^{-1}.$ Clearly, $\pi$ maps each $x_{i}$ onto $\lambda_{i}$, i.e.
\[
\lambda_{i}=A+Bx_{i}.
\]
It is well known (\cite{Blahut}, \cite{MS}) that the extended BCH code $C$ is
invariant under any affine permutation of the locators, so that $\{\lambda
_{i}\}$ is also a locator set in $C_{d-1}.$ Indeed, for any $t\in
\lbrack0,d-3],$ we have an equality%
\begin{align}
\sum\limits_{i=1}^{d-1}\xi_{i}\lambda_{i}^{t}  &  =\sum\limits_{i=1}^{d-1}%
\xi_{i}(A+Bx_{i})^{t}\nonumber\\
&  =\sum\limits_{j=0}^{t}A^{t-j}B^{j}{\binom{t}{j}}\sum\limits_{i=1}^{d-1}%
\xi_{i}x_{i}^{j}=0; \label{main_system1}%
\end{align}
We shall now demonstrate that~(\ref{main_system0}) yields one more equation
\begin{equation}%
\begin{array}
[c]{l}%
\sum\limits_{i=1}^{d-1}\xi_{i}\lambda_{i}^{d-2}=0.
\end{array}
\label{main_system2}%
\end{equation}
Indeed, we use (\ref{Inject_prop}) and~(\ref{Norm}) to obtain
\begin{align}
N_{d-2}\left(  \hat{a}+\lambda_{i}\hat{b}\right)   &  =\left(  \hat{a}+\lambda
_{i}\hat{b}\right)  ^{{q^{(d-3)s}+\ldots+q^{s}+1}}\nonumber\\
&  =\prod\limits_{t=0}^{d-3}\left(  \hat{a}^{q^{ts}}+\lambda_{i}\hat
{b}^{q^{ts}}\right) \label{norm_trans}\\
&  =\sum\limits_{t=0}^{d-2}C_{t}(\hat{a},\hat{b})\lambda_{i}^{t},\nonumber
\end{align}
where $C_{t}$ are some polynomials in $\hat{a}$ and $\hat{b}$. Now the last
equation in (\ref{main_system0}) can be rewritten as%
\[
\sum\limits_{i=1}^{d-1}\xi_{i}\sum\limits_{t=0}^{d-2}C_{t}(\hat{a},\hat
{b})\lambda_{i}^{t}=\sum\limits_{t=0}^{d-2}C_{t}(\hat{a},\hat{b}%
)\sum\limits_{i=1}^{d-1}\xi_{i}\lambda_{i}^{t}=0.
\]
This gives (\ref{main_system2}), due to the two facts:

\begin{itemize}
\item $\sum\limits_{i=1}^{d-1}\xi_{i}\lambda_{i}^{t}=0$ for all $t=0,...,d-3,$
according to ~(\ref{main_system1}).

\item $C_{d-2}(\hat{a},\hat{b})=N_{d-2}(\hat{b})$ and is nonzero, since $b$ is
nonzero and norm is a degree.
\end{itemize}

Equations (\ref{main_system1}) and (\ref{main_system2}) form the linear
system
\[
\begin{array}
[c]{cc}%
\sum\limits_{i=1}^{d-1}\xi_{i}\lambda_{i}^{t}=0,\quad & t=0,...,d-2
\end{array}
\]
in variables $\xi_{i}$ with the Vandermonde matrix $\left(  \lambda_{i}%
^{t}\right)  .$ Recall that $\{\xi_{i}\}$ are nonzero and $\{\lambda_{i}\}$
are pairwise distinct. Therefore these $d-1$ equations hold only if $\xi
_{i}=0$ simultaneously. Thus, our initial assumption that $c$ has weight $d-1$
does not hold. This completes the proof. {\hfill$\Box$}

\smallskip

\textbf{Lemma \refstepcounter{num}\arabic{num} \label{Asymptotics}:} Suppose
char$F_{q}>d-3$; then
\[
c(q,d)\leq(d-3)+1/(d-2).
\]

\textbf{Proof:} We estimate the asymptotic redundancy of the family of codes
presented in Theorem~\ref{Code_family}. $\ $Here $q$ and $d$ are fixed, while
$m>(d-3)!$ runs to infinity over primes. Then
\begin{align}
c(q,d)  &  \leq\lim\limits_{m\rightarrow\infty}\frac{(d-3)m+\left\lceil
m/(d-2)\right\rceil +1}{m}\nonumber\\
&  = (d-3)+1/(d-2). \label{lim}%
\end{align}
The proof is completed. {\hfill$\Box$}

It is obvious that for every $d\geq3$ there exists an infinite family
$\{q_{i}\}$ of growing alphabets such that char$F_{q_{i}}>d-3$. Combining
Lemma ~\ref{Asymptotics} with Corollary~\ref{Alphabet_decrease}, we get
Theorem \ref{Main_theorem}. The proof is completed. {\hfill$\Box$}\smallskip

To conclude, we would like to note that our construction of code $C^{\prime}%
$~(\ref{Final_matrix}) generalizes the construction of nonbinary double
error-correcting codes from Theorem~7 in~\cite{Dumer_95}.

\section{Affine lines}

Before we proceed to the proof of Theorem~\ref{Lines}, let us introduce some
standard concepts and theorems of algebraic geometry. Let $F$ be an
algebraically closed field and $r,t$ be two positive integers. Let
$f_{1},\ldots,f_{r}\in F[x_{1},\ldots,x_{t}].$ For any $x=(a_{1},\ldots
,a_{t})\in F^{t}$, the matrix
\begin{equation}
J_{x}(f_{1},\ldots,f_{r})=\left(
\begin{array}
[c]{ccc}%
\frac{\partial f_{1}}{\partial x_{1}}\bigl|_{x} & \ldots & \frac{\partial
f_{1}}{\partial x_{t}}\bigl|_{x}\\
\vdots & \ddots & \vdots\\
\frac{\partial f_{r}}{\partial x_{1}}\bigl|_{x} & \ldots & \frac{\partial
f_{r}}{\partial x_{t}}\bigl|_{x}%
\end{array}
\right)  \label{Jacobian}%
\end{equation}
is called the \textit{Jacobian} of functions $f_{i}$ at point $x$.

The set $V$ of common roots to the system of equations
\begin{equation}
\left\{
\begin{array}
[c]{ccc}%
f_{1}(x_{1},\ldots,x_{t}) & = & 0,\\
\vdots &  & \\
f_{r}(x_{1},\ldots,x_{t}) & = & 0.
\end{array}
\right.  \label{variety}%
\end{equation}
is called an affine variety. The ideal $I(V)$ is the set of all polynomials
$f\in F[x_{1},\ldots,x_{t}]$ such that $f(x)=0$ for all $x\in V$. One
important characteristic of a variety is its dimension $\dim V$. Dimension of
a non-empty variety is a non-negative integer. Let $x=(a_{1},\ldots,a_{t})\in
V$ be an arbitrary point on $V$. The dimension of a variety $V$ at a point
$x$, denoted $\dim_{x}V$, is the maximum dimension of an irreducible component
of $V$ containing $x$. A point $x\in V$ such that $\dim_{x}V=0$ is called an
\textit{isolated} point.

We shall need the following lemma~(\cite{Kunz}, p.166).

\textbf{Lemma \refstepcounter{num}\arabic{num} \label{Jacobian_dimension}:}
Let $V$ be an affine variety with the ideal
\[
I(V)\subset F[x_{1},\ldots,x_{t}].
\]
Then for any $x=(a_{1},\ldots,a_{t})\in V$ and $f_{1},\ldots,f_{r}\in I(V)$
\[
\mbox{rank}\hspace{0.04cm}J_{x}(f)\leq t-\dim_{x}V.
\]

The next lemma is a corollary to the classical Bezout's
theorem~(\cite{Hartshorne}, p.53).

\textbf{Lemma \refstepcounter{num}\arabic{num} \label{Irreducible_components}%
:} Let $V$ be an affine variety defined by~(\ref{variety}). Then the number of
isolated points on $V$ does not exceed
\[
\prod\limits_{i=1}^{r}\deg f_{i}.
\]

\smallskip

Let $\xi_{1},\ldots,\xi_{t+1}$ be fixed non-zero elements of some finite field
$F_{q}$. Consider a variety $V$ in the algebraic closure of $F_{q}$ defined by
the following system of equations.
\begin{equation}
\left\{
\begin{array}
[c]{cccccccc}%
\xi_{1}x_{1} & + & \ldots & + & \xi_{t}x_{t} & + & \xi_{t+1} & =0,\\
\xi_{1}x_{1}^{2} & + & \ldots & + & \xi_{t}x_{t}^{2} & + & \xi_{t+1} & =0,\\
&  & \vdots &  &  &  &  & \\
\xi_{1}x_{1}^{t} & + & \ldots & + & \xi_{t}x_{t}^{t} & + & \xi_{t+1} & =0.
\end{array}
\right.  \label{Bezout_system}%
\end{equation}
Let $x=(a_{1},\ldots,a_{t})$ be an arbitrary point on $V$. We say that $x$ is
an \textit{interesting} point if $a_{i}\neq a_{j}$ for all $i\neq j$.

\textbf{Lemma \refstepcounter{num}\arabic{num} \label{Jacob_apply}:} Let $V$
be the variety defined by~(\ref{Bezout_system}). Suppose char\hspace
{0.04cm}$F_{q}>t$; then every interesting point on $V$ is isolated.

\textbf{Proof:} Let $x=(a_{1},\ldots,a_{t})$ be an arbitrary interesting point
on $V$. Let $f_{i}(x_{1},\ldots,x_{t})$ denote the left hand side of the
$i$-th equation of~(\ref{Bezout_system}). Consider the Jacobian of $\{f_{i}\}$
at point~$x$.
\[
J_{x}(f_{1},\ldots,f_{t})=\left(
\begin{array}
[c]{ccc}%
\xi_{1} & \ldots & \xi_{t}\\
2\xi_{1}a_{1} & \ldots & 2\xi_{t}a_{t}\\
\vdots & \ddots & \vdots\\
t\xi_{1}a_{1}^{t-1} & \ldots & t\xi_{t}a_{t}^{t-1}%
\end{array}
\right)  .
\]
Thus we have
\[
\det J_{x}(f_{1},\ldots,f_{t})=t!\prod\limits_{i=1}^{t}\xi_{i}\left|
\begin{array}
[c]{ccc}%
1 & \ldots & 1\\
a_{1} & \ldots & a_{t}\\
\vdots & \ddots & \vdots\\
a_{1}^{t-1} & \ldots & a_{t}^{t-1}%
\end{array}
\right|  .
\]
Using standard properties of the Vandermonde determinant and the facts that
${\xi_{i}}$ are non-zero and char\hspace{0.04cm}$F_{q}>t$, we get
\begin{equation}
\mbox{rank}\hspace{0.04cm}J_{x}(f_{1},\ldots,f_{t})=t. \label{rank}%
\end{equation}
It is easy to see that $f_{1},\ldots,f_{t}\in I(V)$. Combining~(\ref{rank})
with Lemma~\ref{Jacobian_dimension}, we obtain $\dim_{x}V=0$. The proof is
completed. {\hfill$\Box$}

\smallskip

\textbf{Lemma \refstepcounter{num}\arabic{num} \label{Rational_solutions}:}
Let $m$ be a prime $m>t!$. Assume char$F_{q}>t$. Let $V$ be the variety
defined by~(\ref{Bezout_system}). Suppose $x\in F_{q^{m}}^{t}$ is an
interesting point on $V$; then $x\in F_{q}^{t}$. In other words, every
interesting point on $V$ that is rational over $F_{q^{m}}$ is rational
over~$F_{q}$.

\textbf{Proof:} Assume the converse. Let $x=(a_{1},\ldots,a_{t})$ be an
interesting point on $V$ such that $x\in F_{q^{m}}^{t}\setminus F_{q}^{t}$.
Consider the following $m$ conjugate points
\[
p_{i}=(a_{1}^{q^{i}},\ldots,a_{t}^{q^{i}}),\mbox{ for all }0\leq i\leq m-1.
\]
Each of the above points is interesting. Since $m$ is a prime, the points are
pairwise distinct. However, according to Lemma~\ref{Jacob_apply} every
interesting point on $V$ is isolated. Thus, we have $m>t!$ isolated point on
$V$. This contradicts Lemma~\ref{Irreducible_components}. {\hfill$\Box$}

\textbf{Remark \refstepcounter{num}\arabic{num} \label{m_form}:} Note that we
can slightly weaken the condition of Lemma~\ref{Rational_solutions} replacing
\[
m\mbox{ prime}\quad\mbox{and}\quad m>t!
\]
with condition: $\forall s\neq1$, $s|m$ implies $s>t!$.

\smallskip

Now we are ready to prove Theorem~\ref{Lines}.

\textbf{Proof:} Assume $C_{d-1}$ is nonempty (this fact will be proven later)
and consider the locator set $X(c)=(x_{1},\ldots,x_{d-1})$ for any $c\in
C_{d-1}$. Recall that $X(c)$ satisfies the first $d-2$ equations
in~(\ref{main_system0}) \ where $\xi_{i}\neq0$ for all $i.$ Consider an affine
permutation $\pi(x)=a+bx$ of the locator set $F_{q^{m}}$ of the code $C$. Let
$a,b\neq0\in F_{q^{m}}$ be such that
\begin{equation}
\pi(x_{d-2})=1\mbox{\ \ and\ \  }\pi(x_{d-1})=0. \label{T}%
\end{equation}
Let $y_{i}$ denote $\pi(x_{i})$. Now we again use the fact that code
$C$~(\ref{BCH_matrix}) is invariant under affine permutations. Therefore the
new locator set $y(c)=(y_{1},\ldots,y_{d-3},1,0)$ satisfies similar equations
\begin{equation}
\left\{  \arraycolsep=0.35em%
\begin{array}
[c]{cccccl}%
\xi_{1} & +\ldots+ & \xi_{d-3} & + & \xi_{d-2} & =-\xi_{d-1},\\
\xi_{1}y_{1} & +\ldots+ & \xi_{d-3}y_{d-3} & + & \xi_{d-2} & =0,\\
\xi_{1}y_{1}^{2} & +\ldots+ & \xi_{d-3}y_{d-3}^{2} & + & \xi_{d-2} & =0,\\
&  & \vdots &  &  & \\
\xi_{1}y_{1}^{d-3} & +\ldots+ & \xi_{d-3}y_{d-3}^{d-3} & + & \xi_{d-2} & =0.
\end{array}
\right.  \label{L_system0}%
\end{equation}

Now we remove the first equation (which does not include variables $y_{i})$
from (\ref{L_system0}), and obtain the system of equations, which is identical
to system~(\ref{Bezout_system}) for $t=d-3$. Recall that $x_{1},\ldots
,x_{d-1}$ are pairwise distinct elements of $F_{q^{m}}$. Therefore
$y_{1},\ldots,y_{d-3},1,0$ are also pairwise distinct. Thus $y_{1}%
,\ldots,y_{d-3}$ is an interesting solution to the above system.

It is straightforward to verify that all the conditions of
Lemma~\ref{Rational_solutions} hold. This yields
\[
y_{i}=a+bx_{i}=\lambda_{i}\in F_{q},\quad\forall i\in\lbrack1,d-1].
\]

Thus, we obtain all locators $x_{i}$ on the affine line%
\[
x_{i}=-\frac{a}{b}+\frac{\lambda_{i}}{b},\quad\lambda_{i}\in F_{q}.
\]
Finally, we prove that $C_{d-1}$ is nonempty. Note that char$F_{q}\geq d-2.$
Also, recall that we consider codes $C_{q}^{m}(d-1)$ with constructive
distance $d-1,$ in which case $q$ does not divide $d-2.$ Thus, we now assume
that $q\geq d-1$. Then we consider (\ref{L_system0}) taking $\xi_{d-1}=1$ and
arbitrarily choosing $d-3$ different locators $y_{1},...,y_{d-3}$ from
$F_{q}\setminus\{0,1\}.$ Obviously, the resulting system of linear equations
has nonzero solution $\xi_{1},\ldots,\xi_{d-2}$. This gives the codeword of
weight $d-1$ and completes the proof of Theorem~\ref{Lines}. {\hfill$\Box$}

\section{Conclusion}

We have constructed an infinite family of nonbinary codes that reduce the
asymptotic redundancy of BCH codes for any given alphabet size $q$ and
distance $d$ if $q\geq d-1$. Families with such a property were earlier known
only for distances $4$, $5,$ and $6$~\cite{Dumer_95}. Even the shortest codes
in our family have very big length $n\approx q^{(d-3)!}$, therefore the
construction is of theoretical interest.

The main question (i.e. the determination of the exact values of $c(q,d)$)
remains open.

\section*{Acknowledgement}

S.~Yekhanin would like to express his deep gratitude to M.~Sudan for
introducing the problem to him and many helpful discussions during this work.
He would also like to thank J.~Kelner for valuable advice.

\end{document}